\documentclass[aps,prb,twocolumn,amsmath,amssymb,floatfix,showpacs]{revtex4}
\usepackage{amsmath}
\usepackage{graphicx}
\usepackage{amssymb}
\usepackage{bm}

\mathchardef\mhyphen="2D
\newcommand{\ttv}{t\mhyphen t^\prime\mhyphen V}

\begin{document}

\preprint{This line only printed with preprint option}

\title{Conductivity of disordered quantum lattice models at infinite 
temperature:\\Many-body localization}

\author{Timothy C. Berkelbach}

\affiliation{Department of Chemistry, Columbia University, 3000 Broadway, New 
York, New York 10027, USA}

\author{David R. Reichman}

\email{drr2103@columbia.edu}

\affiliation{Department of Chemistry, Columbia University, 3000 Broadway, New 
York, New York 10027, USA}

\begin{abstract}
We reinvestigate the behavior of the conductivity of several 
disordered quantum lattice models at infinite temperature using 
exact diagonalization. Contrary to the conclusion drawn in a 
recent investigation of similar quantities in identical systems, 
we find evidence of a localized regime for strong random fields. 
We estimate the location of the 
critical field for the many-body localization transition for the 
random-field XXZ spin chain, and compare our findings with recent 
investigations in related systems.
\end{abstract}


\maketitle

\section{Introduction}
Anderson localization in non-interacting systems is a well 
understood physical process whereby sufficiently strong disorder 
leads to localization of eigenstates and hence insulating 
behavior in systems that would otherwise be conductors.\cite{abr79}
Extensive 
work performed over the last 50 years has established in great 
detail the nature of the Anderson transition in non-interacting 
systems, while much less work has been aimed at elucidating how 
short-ranged interactions modify the simple picture of transport 
that emerges in the non-interacting situation.\cite{lag09} This is somewhat 
surprising, given the fact that attention to the issue of 
interactions already appears in Anderson's classic 1958
paper.\cite{and58}

Recently Basko, Aleiner and Altshuler (BAA) performed a detailed 
diagrammatic analysis demonstrating that weak, short-ranged 
electron-electron interactions generically lead to a finite 
temperature metal-insulator transition in systems that would be 
localized in the absence of interactions.\cite{bas06} In fact the analysis 
and implications of the work of BAA go beyond consideration of 
interacting electrons, suggesting that more general quantum 
entities (e.g. spins, bosons) with local interactions may 
generically fail to thermalize until a threshold energy is 
reached. The notion that such a ``many-body'' localization (MBL)
transition may occur is at odds with the intuition that 
interactions should lead to a finite dc conductivity at all 
finite temperatures in analogy with the mechanism of 
phonon-mediated hopping conductivity. It is also at odds with, 
for example, the analysis of Fleishman and Anderson which argues 
that truly short-ranged interactions are insufficient to induce 
conductivity in an otherwise localized system at any
temperature.\cite{fle80}

The work of BAA has motivated more recent investigations of the 
possibility of a finite temperature transition from a localized 
(non-ergodic) phase to a delocalized phase where interactions 
afford thermalization of the system. These works, both analytical 
and numerical, have reached somewhat conflicting
conclusions.\cite{oga07,zni08,kar09,mue09,iof09} In 
this work, we reconsider the analysis of Karahalios et al.\cite{kar09} Via 
examination of the conductivity, these authors concluded that, in 
general, finite temperature systems of one-dimensional 
interacting spins are always conducting, thus contradicting the 
claim of BAA.

\section{Imaginary frequency conductivity}

Here, as in previous work, we make use of the 
important observation of Oganesyan and Huse that the many-body 
localization transition may be probed at infinite temperature by 
varying the disorder strength.\cite{oga07} This simplifies the problem by 
reducing the number of control parameters that may be varied to 
tune the system from a delocalized to a localized phase. As in 
the work of Karahalios et al., we examine one-dimensional spin 
chains via exact diagonalization, calculating the 
conductivity via the Kubo formula. An important conclusion of our 
work is that sufficient care needs to be exercised in the 
interpretation of the zero frequency conductivity as a function 
of disorder strength and level broadening.

For a finite system of length $L$ at $T \rightarrow \infty$,
the Kubo formula for the conductivity is given by
\begin{equation}
	\sigma(\omega) = \frac{\beta}{L} \lim_{\eta\to 0} 
		\int_0^\infty e^{i(\omega + i\eta)t} \langle j(t) j(0) \rangle dt,
\end{equation}
where $\beta=1/k_BT$, $j(t)$ is the current operator at time $t$,
and $\eta$ can be thought of as both a numerical tool for
convergence as well as a phenomenological level broadening for
discrete spectra.\cite{imr02} The real part of the conductivity may be
decomposed as $\sigma^\prime(\omega)=D\delta(\omega) + 
\sigma_{\rm reg}(\omega)$, where the Drude weight $D$ measures
purely ballistic conduction and arises due to pairs of degenerate
states connected by the current operator.  However, in the systems
studied here, all level degeneracies are lifted in the presence of disorder,
and conductivity is purely diffusive.  Thus we may take as our definition of
the dc conductivity $\sigma_{\rm dc} = \sigma(\omega\rightarrow0)$ without
concern for the Drude contribution. By employing the spectral representation of
the Hamiltonian we arrive at the `imaginary frequency' dc conductivity,
\begin{equation}
	\sigma(i\eta) = \frac{\beta}{ZL} \sum_{m,n} 
			|\langle m|j|n\rangle|^2 
			\frac{\eta}{\eta^2 + (\omega_{nm})^2},
\end{equation}
where $H|n\rangle = E_n|n\rangle$, $\omega_{nm}=E_n-E_m$, and $Z$ is
the partition function. 

The authors of Ref.[8] calculate the full frequency-dependent
(ac) conductivity spectrum using a level-broadening binning procedure
and draw conclusions regarding dc conductivity based on the $\omega
\rightarrow 0$ behavior.  However, as we will show, all finite-sized
systems with level broadening will exhibit a non-zero dc conductivity,
and thus such an analysis is inconclusive.  Rather, it is the conductivity's
{\em dependence} on this level broadening which allows one to draw
conclusions regarding conducting and insulating behaviors in the thermodynamic
limit.

As discussed by 
Thouless and Kirkpatrick,\cite{tho81} finite systems
should display a simple asymptotic behavior for the 
dc conductivity that scales as $\eta$ for small 
$\eta$ and $\eta^{-1}$ for large $\eta$. The distinction between 
conductor and insulator manifests in the behavior between 
these asymptotic regimes. In particular, we expect that a system 
with insulating behavior will exhibit an imaginary frequency dc
conductivity with well resolved $\eta$ and $\eta^{-1}$ 
regimes separated by a simple maximum. Finite sized systems
expected to behave as conductors in the $L \rightarrow \infty$ 
limit exhibit a broad crossover between these regimes, 
with a plateau signifying the onset of a true dc 
conductivity.\cite{tho81,imr02} Although
the dc conductivity is well defined only in the thermodynamic
limit (specifically, $L\rightarrow\infty$, 
then $\eta\rightarrow0$), we see evidence for
dc conductivity manifesting itself even at the small system sizes
accessible by exact diagonalization. 

It should be pointed out that although the analysis employed here
was originally developed for non-interacting systems, our results
empirically show that it is equally applicable to interacting ones,
by replacing the single-particle levels with many-body levels.
Specifically, we locate an insulating regime in which 
$\sigma(i\eta)\propto \eta$ when $\eta$ is less than the level
spacing in the many-body localization volume, which does
not scale with the size of the
system.  This behavior is to be contrasted with an observed
metallic regime, which has $\sigma(i\eta) \propto \eta$ as
long as $\eta$ is less than the
many-body level spacing in the system volume
-- a spacing which vanishes in the thermodynamic limit yielding
a dc conductivity plateau at small to intermediate $\eta$.

\section{Quantum Lattice Models}

We study two quantum lattice models in the presence of disorder:
the $XXZ$ spin chain and the $\ttv$ model of spinless
fermions, originally studied in
its disordered form by Oganesyan and Huse\cite{oga07} and more
recently by Monthus and Garel.\cite{mon10} The disordered
$XXZ$ chain is given by the Hamiltonian
\begin{align}
H_{XXZ} = & J \sum_{j=1}^{L} \left[ S^x_{j} S^x_{j+1} + S^y_{j} S^y_{j+1}
		+ \Delta S^z_{j} S^z_{j+1}\right] \\
	&+\sum_{j=1}^{L} w_j S^z_j, \notag
\end{align}
where we choose the random fields
$w_j$ uniformly from $[-W,W]$.  The current operator for the
$XXZ$ chain is given by
\begin{equation}
j_{XXZ} = J\sum_{j=1}^{L} \left[ S^x_{j} S^y_{j+1} 
				- S^y_{j} S^x_{j+1} \right].
\end{equation}

\begin{figure}[t]
   \includegraphics[scale=0.33]{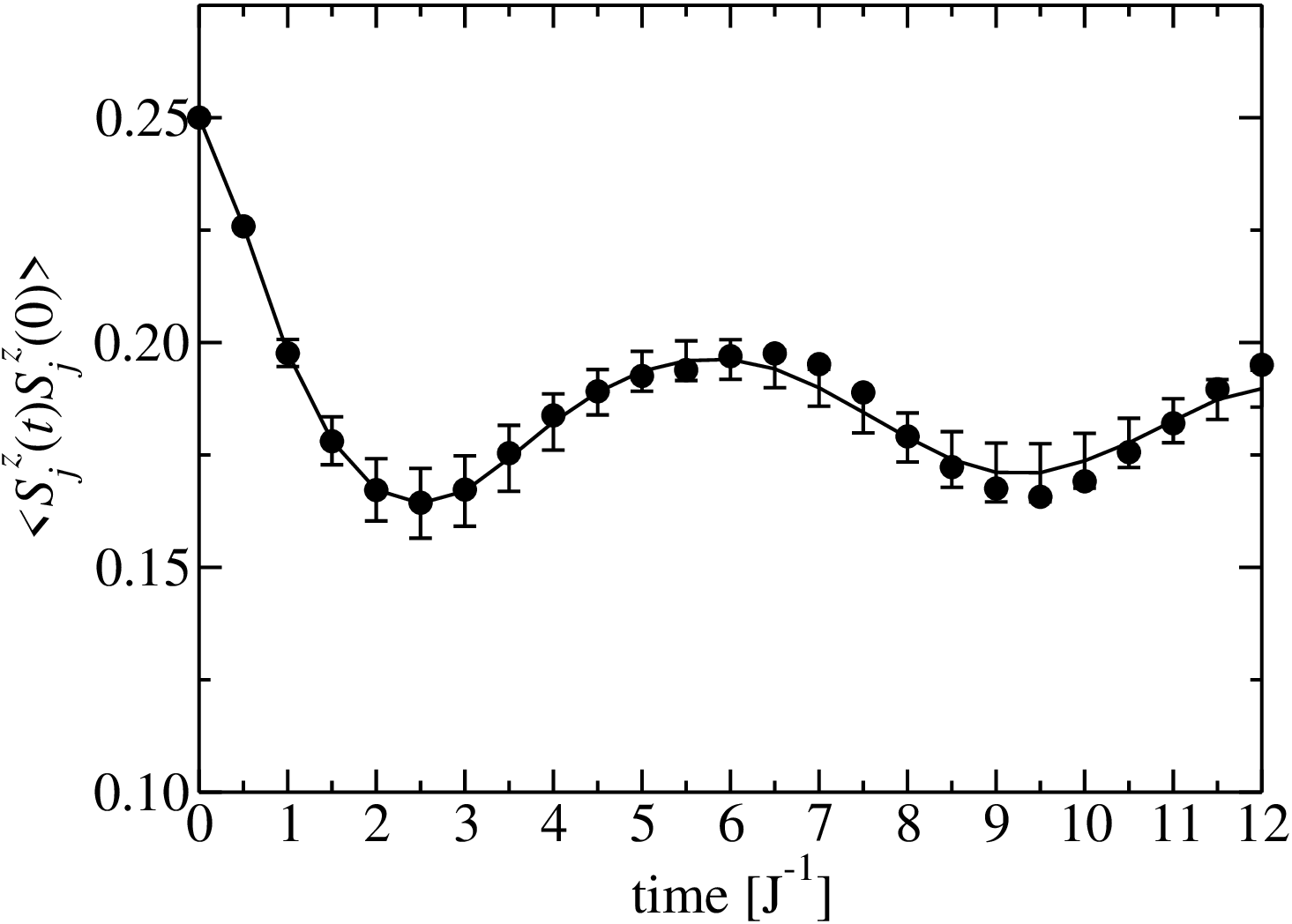}
   \caption{Local spin-spin correlation function,
	$\langle S^z_j(t)S^z_j(0)\rangle$ for the disordered
	$XXZ$ chain with $W=5$.  Results are shown for the
	non-interacting case, $\Delta=0.0$ (filled circles) and
	for the interacting case, $\Delta=0.5$ (solid line).
	Error bars for the non-interacting case (not shown for
	clarity) are of the same order as those shown for the
	interacting case.
	}
   \label{fig:spin}
\end{figure}

The disordered $\ttv$ model is described by the Hamiltonian
\begin{align}
H_{\ttv} = \sum_{j=1}^{L} &\bigg[ -t\left(c_j^\dagger c_{j+1} + c_{j+1}^\dagger c_j\right) \\
		&- t'\left(c_j^\dagger c_{j+2} + c_{j+2}^\dagger c_j\right) \notag\\
		&+ V \left(n_j-\frac{1}{2}\right) \left(n_{j+1}-\frac{1}{2}\right) + w_j n_j \bigg], \notag
\end{align}
where, following Oganesyan and Huse, the random on-site energies $w_j$
are chosen from a Gaussian distribution with mean $0$ and
variance $W^2$.  The $\ttv$ model's current operator is
\begin{align}
j_{\ttv} = i\sum_{j=1}^{L} &\left[
	t\left(c_j^\dagger c_{j+1} - c_{j+1}^\dagger c_j\right) \right. \\
	&+ \left. 2t^\prime\left(c_j^\dagger c_{j+2} 
		- c_{j+2}^\dagger c_j\right) \right]. \notag
\end{align}

\begin{figure}[t]
   \includegraphics[scale=0.34]{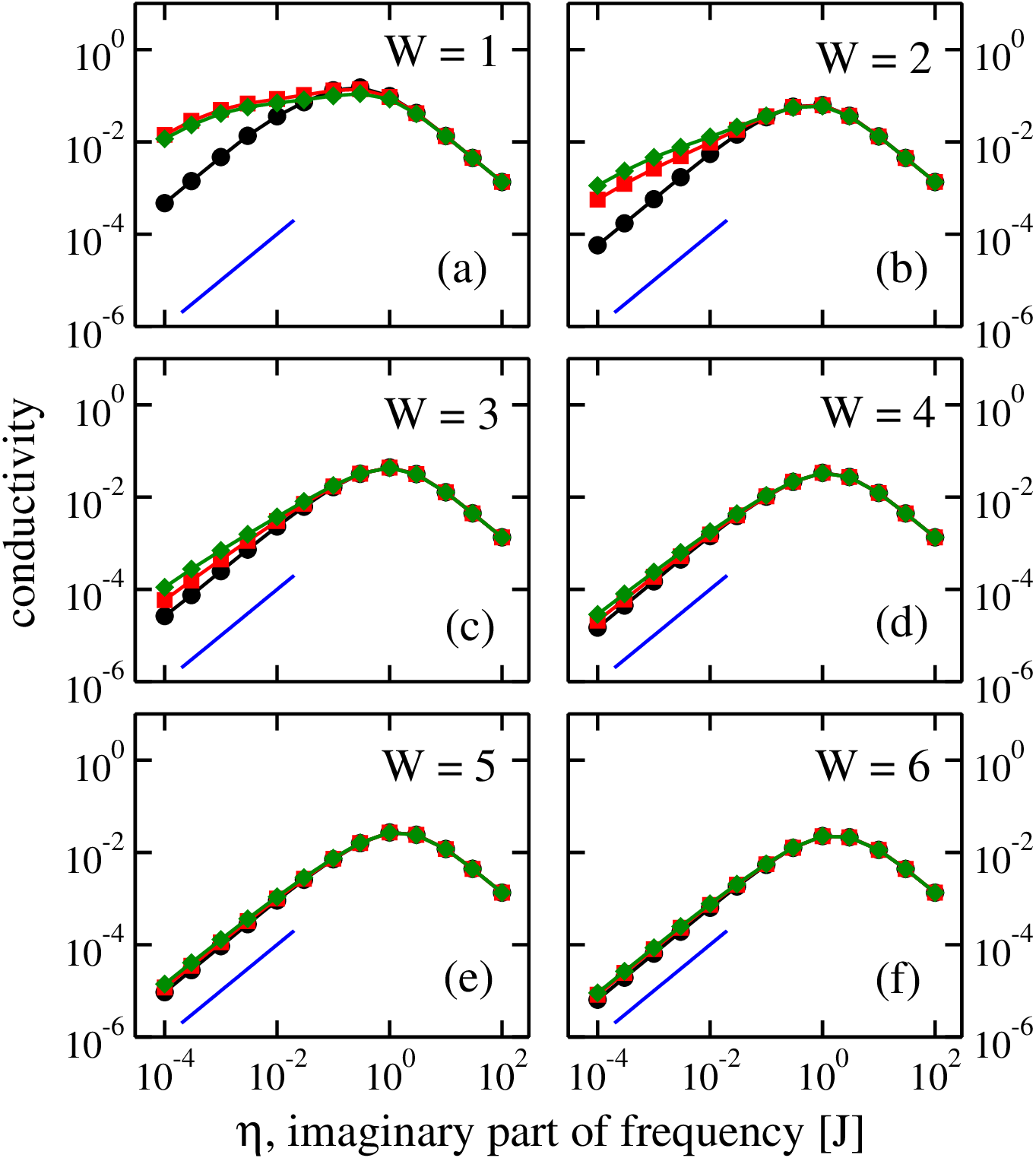}
   \caption{The dc conductivity as a function of the imaginary frequency, $\eta$,
	for the disordered $XXZ$ chain ($J=1$) with
	$W = 1-6$ [(a)$-$(f)] for the non-interacting case, $\Delta=0.0$
	(black line, circles) and for two interacting cases,
	$\Delta=0.5$ (red line, squares) and $\Delta=1.0$ (green
	line, diamonds). Error bars are smaller than
	the symbols. The blue line shows linear slope.
	}
   \label{fig:xxz}
\end{figure}

In what follows, we restrict our Hilbert space to $S^z_{tot}=0$ 
for the $XXZ$ chain and to half-filling for the $\ttv$ model.
We have explored other alternatives and find our results to be
qualitatively similar. All results are presented for $L=14$
with periodic boundary conditions, 
although we have studied systems as large as $L=16$ and find
our conclusions unaltered.
We average over $N_r=100$ independent realizations
of disorder and calculate error bars as the standard deviation of the
mean, $\pm \sigma/\sqrt{N_r}$, where $\sigma$ (not to be confused with
the conductivity) is the standard deviation across disorder realizations.
Our results appear to be converged, although for 
such a small range of system sizes one should view such 
statements with care. This is especially true for smaller values 
of disorder, where finite localization lengths, if they exist, 
can clearly be larger than the system sizes accessible from exact 
diagonalization.

The aforementioned claim of Karahalios et al. is rather surprising in light of 
the fact that an earlier tDMRG calculation performed on strongly disordered
spin chains found evidence of a localized phase, at least when 
viewed from the standpoint of the local spin-spin correlation 
function.\cite{zni08} In Fig.~\ref{fig:spin} we reproduce this result 
(for a smaller 
system and the shorter time scales accessible in exact 
diagonalization) and compare it with dynamics in the 
non-interacting system which is known to be localized.
It is clear that, for the interacting system, the local spin-spin 
correlation exhibits 
quantitatively similar relaxation compared to the localized system and 
shows no sign of decay from a plateau (the analog of the 
Edwards-Anderson parameter), indicative of a glassy phase.

\begin{figure}[t]
   \includegraphics[scale=0.33]{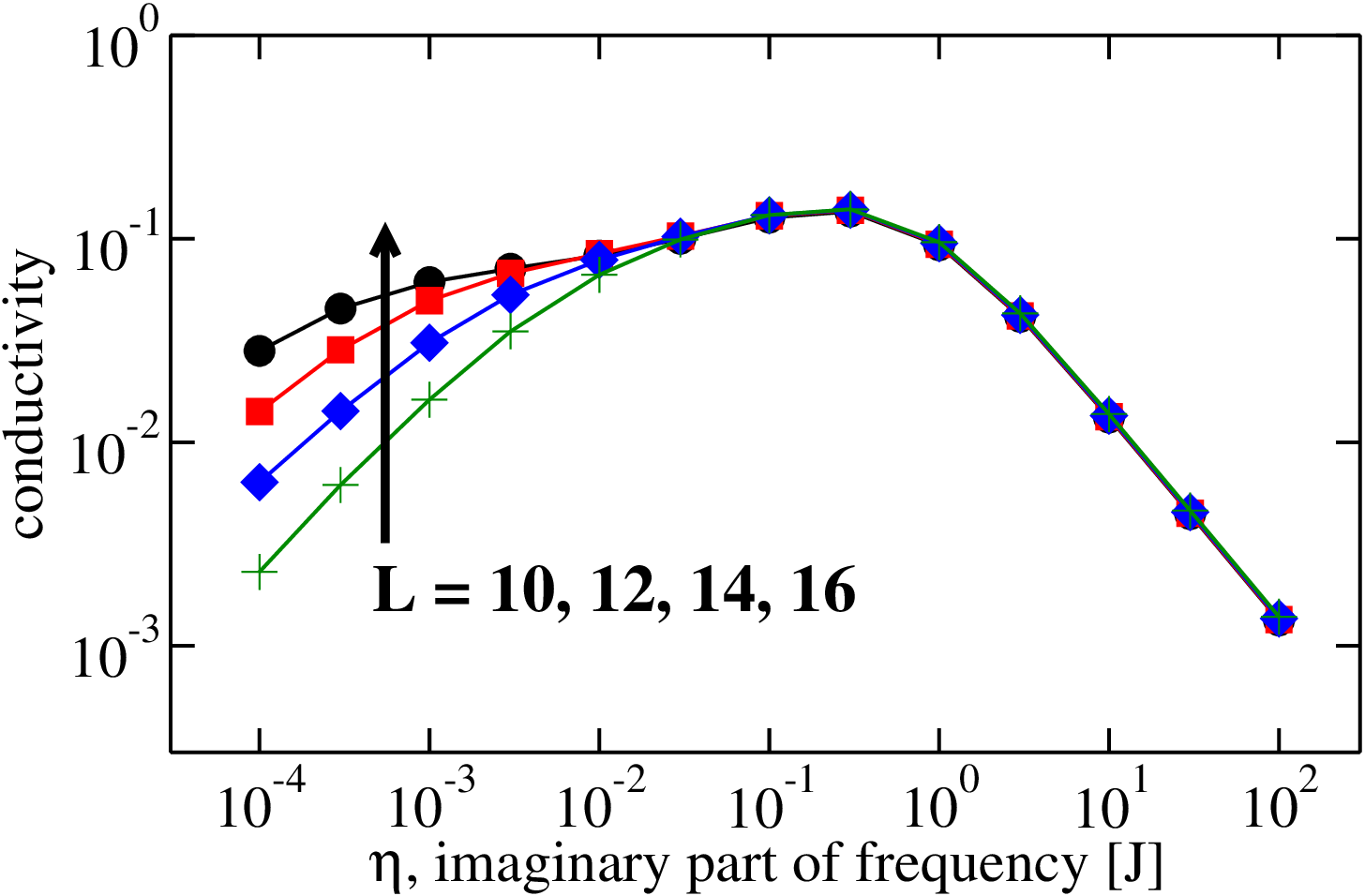}
   \caption{System size dependence of the dc conductivity as
	a function of the imaginary frequency, $\eta$,
	for the disordered $XXZ$ chain ($J=1$) with
	$\Delta = 0.5$ and $W = 1$.  Conductivities are presented
	for $L = 10$ (green plusses), 12 (blue diamonds),
	14 (red squares), and 16 (black circles), showing the development
	of the dc conductivity plateau in the $L \rightarrow \infty$ limit.
	}
   \label{fig:scaling}
\end{figure}

\section{Results and the many-body localization transition}

We turn next to an investigation of the conductivity in the above
disordered lattice models. As discussed above, it is useful
to compare the $\eta$ dependence of the dc conductivity in the interacting
models directly with their noninteracting counterparts to set 
a baseline for localization. In Fig.~\ref{fig:xxz} we present results for two 
different values of the anisotropy $\Delta$ in the $XXZ$ spin chain. 
We extend the results of Karahalios et al., who examine disorder strengths
only as high as $W=1$, by performing our calculations up to $W=6$.
Clearly, by 
$W=4$ in both interacting cases the conductivity curves are essentially
indistinguishable from the non-interacting localized case. Thus, 
at least with regard to exact diagonalization on these system 
sizes and based solely on examination of the conductivity, the 
interacting behavior is identical to the non-interacting, localized behavior. 
Using this condition, we can place an upper limit, 
$W_{c}\lesssim 4$ in both cases $\Delta=0.5$ and
$\Delta=1.0$.\footnote{It should be noted that it is possible that a residual
conductivity, impossible to resolve via an exact diagonalization study
of $\sigma_{dc}$ on small systems, exists. We see no evidence for
finite size effects here, though we are restricted to very small $L$.}
It should be noticed as well that already at 
$W=1$ there is significant structure in $\sigma(i\eta)$,
exhibiting a nearly flat region in between the small and large $\eta$ regimes.
This suggests that these interacting data are 
in a conducting regime, although care must be used because this 
also might be an indication of localization behavior on length 
scales larger than we can access via exact diagonalization.
One should also note that this conducting behavior is fully consistent
with the conclusions of Karahalios et al. at $W = 1.0$. 
With the above caveats, we can place the critical value of $W$ for a 
many-body localization transition in the range
$3 < W_{c} < 4$.\footnote{After this work was completed we became aware of the work of
Pal and Huse (\eprint{arXiv:1003.2613v1}), who also investigate
the $XXZ$ chain with $\Delta=1$. Their conclusions are quantitatively
consistent with our results.  It should be noted that
their work goes beyond that presented here by connecting the MBL
transition to the infinite randomness universality class.} 
It is thus unsurprising that Karahalios et al. found no evidence for the MBL transition,
as we have shown that it occurs at disorder strengths larger than those
investigated in Ref.[8].
We have additionally analyzed adjacent many-body level-spacings (not shown here),
whose crossover from the Gaussian Orthogonal Ensemble to Poisson
statistics\cite{shk93,hof93,hof94}
occurs at this same critical strength of disorder, confirming the robustness
of our approach.

\begin{figure}[t]
   \includegraphics[scale=0.31]{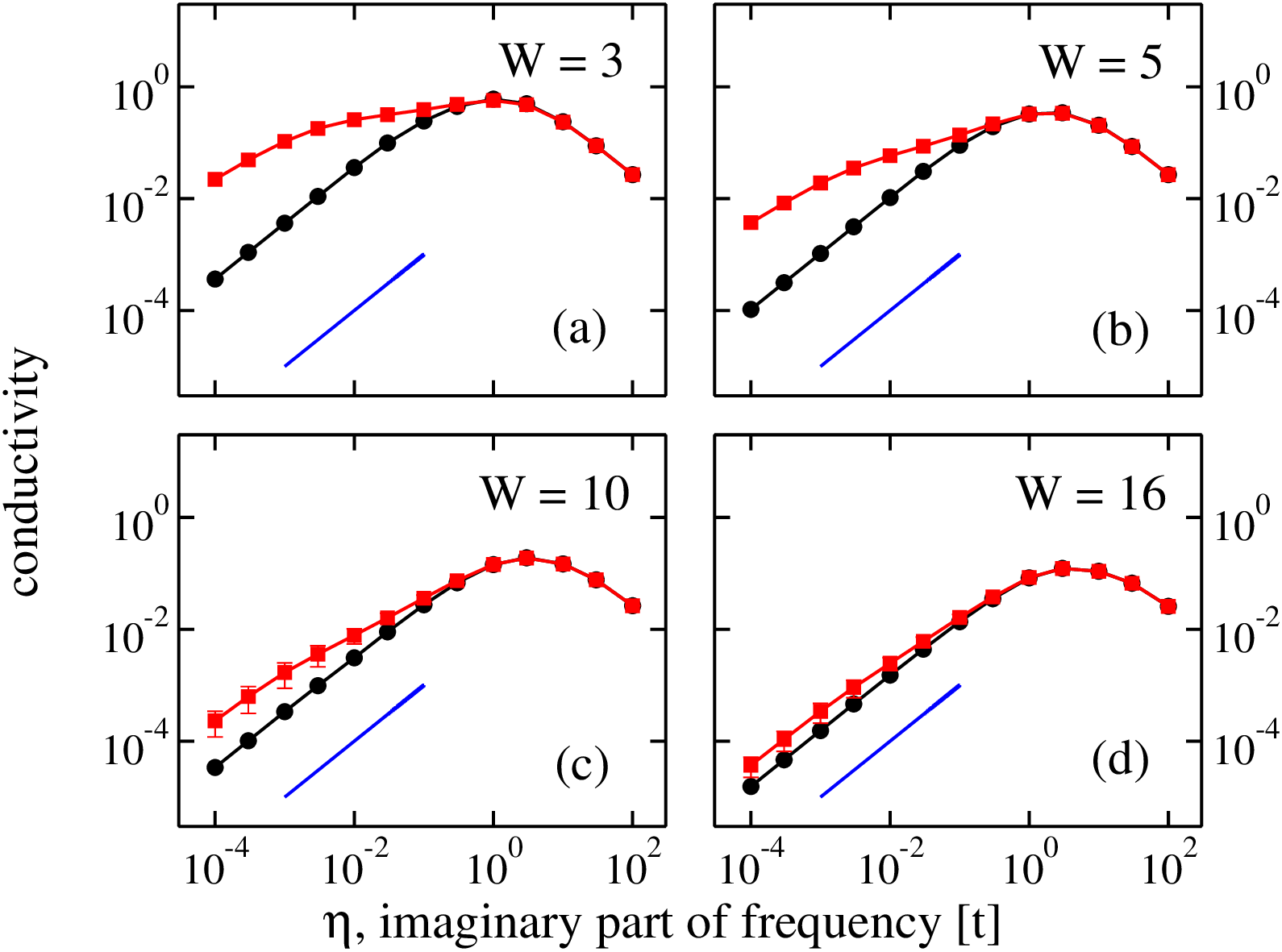}
   \caption{The same as in Fig.\ref{fig:xxz} but for the 
	disordered $\ttv$ model 
	($t=t^\prime=1$) with $W = 3$ (a), $W = 5$ (b), $W=10$ (c),
	and $W = 16$ (d),
	for the non-interacting case, $V=0$
	(black line, circles) and for the interacting case,
	$V=2$ (red line, squares). Where not shown, error
	bars are smaller than the symbols. The blue line shows
	linear slope.
	}
   \label{fig:ttv}
\end{figure}

In order to confirm our expectations of a conducting phase in the
thermodynamic limit, $L \rightarrow \infty$, we have examined the effects 
of system size on the conductivity of the interacting $XXZ$
spin chain with $\Delta = 0.5$.  We focus on the disorder strength
$W = 1$ because of its apparent conducting behavior in Fig.~\ref{fig:xxz}.
Furthermore, one is in danger of approaching localization lengths
equal to the size of the sytem for values of disorder much smaller 
than this.  We show in Fig.~\ref{fig:scaling} the conductivities for
system sizes $L=$ 10, 12, 14, and 16.  Results for $L=16$ are shown
for $N_r = 50$ realizations of disorder.  Clearly, the plateau becomes
more resolved for larger system sizes, strongly suggesting a non-zero
dc conductivity in the thermodynamic limit.  Furthermore, the plateau's growth
extends toward smaller $\eta$ in agreement with the many-body level
spacing discussion above.

We have also examined the behavior of the 
disordered $\ttv$ model of Oganesyan and Huse, the results of which
are presented in Fig.~\ref{fig:ttv}. The behavior is qualitatively
the same as above, suggesting the existence of a MBL transition in
this model as well. 
However, it should be pointed out that the crossover
between apparent conducting and insulating behavior in
$\sigma(i\eta)$ is significantly broader in the $\ttv$
model than in the $XXZ$ systems, making a prediction of the critical
disorder strength difficult.  At larger system sizes
this crossover should become more abrupt, but unfortunately such
sizes are beyond the reach of exact diagonalization.
Despite the above difficulties, we would expect,
based on the same means of analysis presented above, that the 
critical value of $W$ in this model is higher than 
the $W_{c}\approx5$ range found in the real-space renormalization group 
calculation of Monthus and Garel. 
Although, it is not clear when studying finite
systems that different quantities, such as those investigated here and by Monthus
and Garel, should behave in a similar manner. However, our result does strongly
suggest conducting behavior at $W\approx5$. 

\section{Conclusions}

To summarize, we have carried out a systematic investigation of the 
diffusive conductivities of two
common disordered quantum lattice models: the $XXZ$ spin
chain and the $\ttv$ model of spinless fermions. 
We find that by studying the behavior of $\sigma(i\eta)$ we
can place reasonable bounds on the location of the MBL transition.
By examining disorder strengths higher than those explored in previous
works, we find for the disordered $XXZ$ chain (both $\Delta=0.5$ and
$\Delta=1.0$) that the finite size conductivity extrapolates
quantitatively to the non-interacting values by $W\approx4$.  Our
results are also qualitatively consistent with the recent work of
Monthus and Garel, but we would expect based on finite size studies
of the conductivity, that the transition occurs at a disorder value
larger than that found by their numerical renormalization group procedure.

\begin{acknowledgments}
We thank G. Biroli, V. Oganesyan, A. Millis, and F. Zamponi
for useful discussions. We would also like to thank the NSF
for financial support under Grant No. CHE-0719089.
\end{acknowledgments}

\bibliographystyle{apsrev}
\bibliography{mbl}

\end{document}